\documentclass[pra,twocolumn,superscriptaddress]{revtex4-2}

\usepackage[utf8]{inputenc}
\usepackage[T1]{fontenc}
\usepackage{graphics,amsfonts}
\usepackage{epstopdf}
\usepackage[pdftex]{graphicx}
\usepackage[cmex10]{amsmath} 
\usepackage{enumitem}
\usepackage[top=1.5cm, bottom=2cm, right=1.6cm,left=1.6cm]{geometry}
\usepackage{indentfirst}
\usepackage{ulem}
\usepackage[dvipsnames]{xcolor}
\usepackage[caption=false]{subfig}
\usepackage{hyperref}

\graphicspath{{./}{./Images/}}

\DeclareMathOperator{\vac}{vac}

\begin{document}
\title{Security bounds for decoy-state QKD with arbitrary photon-number statistics}
	
\author{Giulio Foletto}
\thanks{These authors contributed equally to this work.}
\affiliation{Dipartimento di Ingegneria dell'Informazione, Universit\`a degli Studi di Padova, via Gradenigo 6B, IT-35131 Padova, Italy}

\author{Francesco Picciariello}
\thanks{These authors contributed equally to this work.}
\affiliation{Dipartimento di Ingegneria dell'Informazione, Universit\`a degli Studi di Padova, via Gradenigo 6B, IT-35131 Padova, Italy}

\author{Costantino Agnesi}
\affiliation{Dipartimento di Ingegneria dell'Informazione, Universit\`a degli Studi di Padova, via Gradenigo 6B, IT-35131 Padova, Italy}

\author{Paolo Villoresi}
\affiliation{Dipartimento di Ingegneria dell'Informazione, Universit\`a degli Studi di Padova, via Gradenigo 6B, IT-35131 Padova, Italy}
\affiliation{Padua Quantum Technologies Research Center, Universit\`a degli Studi di Padova, via Gradenigo 6B, IT-35131 Padova, Italy}

\author{Giuseppe Vallone}
\email{vallone@dei.unipd.it}
\affiliation{Dipartimento di Ingegneria dell'Informazione, Universit\`a degli Studi di Padova, via Gradenigo 6B, IT-35131 Padova, Italy}
\affiliation{Dipartimento di Fisica e Astronomia, Universit\`a degli Studi di Padova, via Marzolo 8, IT-35131 Padova, Italy}
\affiliation{Padua Quantum Technologies Research Center, Universit\`a degli Studi di Padova, via Gradenigo 6B, IT-35131 Padova, Italy}

\begin{abstract}
The decoy-state method is a standard enhancement to quantum key distribution (QKD) protocols that has enabled countless QKD experiments with inexpensive light sources. 
However, new technological advancements might require further theoretical study of this technique.
In particular, the decoy-state method is typically described under the assumption of a Poisson statistical distribution for the number of photons in each QKD pulse.
This is a practical choice, because prepare-and-measure QKD is often implemented with attenuated lasers, which produce exactly this distribution.
However, sources that do not meet this assumption are not guaranteed to be compatible with decoy states.
In this work, we provide security bounds for decoy-state QKD using a source with an arbitrary photon emission statistic.
We consider both the asymptotic limit of infinite key and the finite-size scenario, and evaluate two common decoy-state schemes: the vacuum+weak and one-decoy protocols.
We numerically evaluate the performance of the bounds, comparing three realistic statistical distributions (Poisson, thermal, binomial), showing that they are all viable options for QKD.
\end{abstract}

\maketitle
	
\section{Introduction}\label{sec:intro}
Since its first proposal, the decoy-state method \cite{Hwang2003} has had a major impact on the practicality of quantum key distribution (QKD), having eased many of its technical achievements \cite{Liao2017_Sat, Boaron2018} and, recently, its implementations in field trials \cite{Dynes2019, Avesani:21}.
By countering the photon-number-splitting (PNS) attack \cite{PhysRevLett.85.1330}, it has enabled the use of practical light sources in QKD and it has been object of many studies including implementation proposals and security proofs.
X.~Ma \textit{et al.} presented the first rigorous analysis \cite{Ma2005}, showing that the vacuum+weak scheme is asymptotically optimal. Afterwards, more specific analyses have been made for finite-key realizations of the protocol \cite{Lim2014}, and it has been found that the one-decoy scheme achieves higher secret key rates in some experimental conditions \cite{Rusca2018}.

Most decoy-state analyses intrinsically assume a Poisson distribution for the number of photons in each pulse emitted by the source, since it is the one produced with strongly attenuated lasers. 
However, as the research on innovative light sources progresses, a comprehensive analysis of the decoy-state method beyond the Poisson case must be made. 
Investigations restricted to specific technologies exist \cite{Wang2007,Mauerer2007,Adachi2007,Curty2010,Huang2018}, but a more general study that makes minimal assumptions on the statistical distribution and includes finite-size effects is still needed.
This would enable the use of non-laser sources, such as LEDs \cite{Duligall_2006, Xia:2019} and quantum dots \cite{Heindel_2012, Takemoto2015}, which can still emit multi-photon pulses and are vulnerable to PNS and would provide rigorous bounds to the secure key rate achievable with such sources.

In this work, after introducing the protocol and the formalism (Sec. \ref{sec:decoy}), we provide bounds needed to realize decoy-state QKD with an arbitrary photon-statistic source.
We start from the asymptotic limit of infinitely long key (Sec. \ref{sec:asympt}), and then explain the transition to the finite-key scenario (Sec. \ref{sec:finite}).
In both cases, we consider the vacuum+weak and one-decoy schemes.
Finally, we numerically compare the performance of three distributions (Poisson, thermal, and binomial) showing that they are all viable options for QKD (Sec. \ref{sec:simulations}).

\section{Protocol description}\label{sec:decoy}

The generic formulation of the BB84 protocol with active decoy states considers two parties, a transmitter (Alice) and a receiver (Bob), that share a quantum and a classical channel, through which they can communicate.
Alice chooses a probability distribution for the number of photons in each optical pulse she prepares \cite{Hwang2003, Lo2005,Wang2005,Ma2005}.
The selection is random and follows discrete probabilities $p_k$.
We label $\mathcal D_{k}$ the distributions at her disposal, where $k\in\{\mu, \nu,\ldots\}$ is an identifier, and $P_{i|k}$ the probability of producing an $i$-photon pulse using distribution $\mathcal{D}_k$.
Usually these distributions have the same form and differ only for their parameters.
In the most common case, these are the Poisson distributions that characterize the number of photons in coherent pulses and differ only by their mean values.

For each pulse, Alice chooses between two bases to encode the qubits, while on the other side, Bob chooses between the same bases to measure them.
We work in the context of the \textit{efficient BB84} protocol, which allows us to carry out the analysis separately for the two bases \cite{Lo2005a}.
This choice is not necessary and our results can easily be translated to the standard BB84 case.
However, the efficient variant offers better performance and simplifies the implementation by relaxing some requirements of symmetry in the states.
In this scheme, one of the two bases is chosen much more often and used to generate the key, while the other is used to evaluate the incidence of attacks.
We will label $Z$ the former and $X$ the latter, and we will use the symbol $b\in\{X,Z\}$ for the generic basis, when specifying one is not required.

After a large enough number of states has been detected by Bob, Alice publishes the entire sequence of her choices of bases and probability distributions.
In this way, Alice and Bob together can directly measure the \textit{gains} $Q_{b,k}$, i.e., the conditional frequencies of Bob observing a detection event given that Alice chose the distribution $\mathcal{D}_k$ and the bases chosen by Alice and by Bob are both $b$.

By performing classical error correction procedures, the two can also measure the conditional error rates $E_{b,k}$ for each configuration, that are the conditional frequencies of finding a mismatch between Alice's encoded symbol and Bob's decoded one, given that Bob detected something and that the chosen distribution and bases were $\mathcal{D}_k$ and $b$.

The goal of the procedure is to find the \textit{yields} $Y_{b,i}$ of the photon numbers and the error rate per photon number $e_{b,i}$, that are respectively the conditional probabilities of detections and mismatches given that Alice sent a pulse with a specific number of photons $i$ and the chosen basis was $b$.

These two values are important because they allow to estimate the number of detection events that originate from insecure multi-photon pulses and the amount of information that might have leaked to an eavesdropper Eve. 

We note that, contrarily to the gains $Q_{b,k}$ and the error rates of each distribution $E_{b,k}$, the yields and the error rates per photon number are not directly available because Alice cannot control the photon number, but only its probability distribution.
However, these values can still be upper and lower bounded from experimental quantities and we will see that only $Y_{b,0}$, $Y_{b,1}$ and $e_{b,1}$ are needed for our purpose, because they are the only ones that correspond to secure pulses.

\section{Asymptotic security bounds}\label{sec:asympt}
We present a way to relate all these quantities that does not restrict the probability distributions $P_{i|k}$ to a specific form, but it can be applied to generic distributions that satisfy a simple to verify condition, see Eq. \eqref{eq:first_restriction}.
We start from the ideal limit of infinitely long key blocks, which allows us to neglect statistical errors.

We consider two common decoy-state configurations: the \textit{vacuum + weak} protocol, which requires three probability distributions, $P_{i|\mu}$, $P_{i|\nu}$ and $P_{i|\vac}$, with $P_{0|\vac}=1, P_{i|\vac}=0, \forall i\neq 0$, and the \textit{one-decoy} protocol, which only uses the two distributions $P_{i|\mu}$ and $P_{i|\nu}$ \cite{Ma2005,Rusca2018}. Without loss of generality, we assume $\mu > \nu > 0$.

To apply our method to the \textit{vacuum + weak} protocol, the probability distributions $P_{i|\mu}$ and $P_{i|\nu}$ should satisfy the following condition:
\begin{equation}
    \frac{P_{1|\nu}}{P_{1|\mu}} > \alpha := \max_{i\geq 2}\frac{P_{i|\nu}}{P_{i|\mu}}.
    \label{eq:first_restriction}
\end{equation}

We start by considering that
\begin{align}
    \langle Q_{b,k}\rangle &= \sum_i P_{i|k} Y_{b,i},\label{eq:ExpValGain}\\ 
    \langle E_{b,k} Q_{b,k}\rangle &=\sum_i P_{i|k} e_{b,i} Y_{b,i},\label{eq:ExpValError}
\end{align}
where $\langle\cdot\rangle$ labels the expectation value of a quantity.
In the asymptotic limit, these expectation values are readily available: $\langle Q_{b,k}\rangle = Q_{b,k}$ and $\langle E_{b,k} Q_{b,k}\rangle = E_{b,k} Q_{b,k}$.

Combining Eqs. \eqref{eq:first_restriction}, \eqref{eq:ExpValGain}, \eqref{eq:ExpValError} and  considering that $e_{b,0} = \frac12$. we can find the following bounds:
\begin{align}
    Y_{b,1} \geq Y_{b,1}^L &= \frac{Q_{b,\nu}-\alpha Q_{b,\mu}-(P_{0|\nu}-\alpha P_{0|\mu})Y_{b,0}}{P_{1|\nu}-\alpha P_{1|\mu}},\label{eq:generic_y1l} \\
    e_{b,1}\leq e_{b,1}^U &= \frac{E_{b,\nu} Q_{b,\nu}-P_{0|\nu}Y_{b,0}/2}{P_{1|\nu}Y_{b,1}^L}.
    \label{eq:generic_e1u}
\end{align}

The only missing quantity in the right-hand sides is $Y_{b,0}$. In the vacuum + weak case, it is directly available as $Y_{b,0} = Q_{b,\vac}$.

For the one-decoy protocol we need further condition on the distributions, namely:
\begin{equation}
    \frac{P_{0|\nu}}{P_{0|\mu}}>\frac{P_{1|\nu}}{P_{1|\mu}}.
    \label{eq:second_restriction}
\end{equation}
Because of it, we can find a lower and an upper bound:
\begin{align}
    Y_{b,0} \geq Y_{b,0}^L &= \frac{P_{1|\mu}Q_{b,\nu}-P_{1|\nu}Q_{b,\mu}}{P_{1|\mu}P_{0|\nu}-P_{1|\nu}P_{0|\mu}}, \label{eq:generic_y0l} \\
    Y_{b,0}\leq Y_{b,0}^U &= \frac{2E_{b,\mu} Q_{b,\mu}}{P_{0|\mu}}.
\end{align}
Given that Eq. \eqref{eq:second_restriction} guarantees that $P_{0|\nu}-\alpha P_{0|\mu}>0$, the upper bound $Y_{b,0}^U$ should be inserted into \eqref{eq:generic_y1l}.
Although Eq. \eqref{eq:generic_e1u} could be completed with $Y_{b,0}^L$, we can find a tighter bound thanks to Eq. \eqref{eq:second_restriction}:
\begin{equation}
    e_{b,1}\leq e_{b,1}^U=\frac{P_{0|\nu}E_{b,\mu}Q_{b,\mu} - P_{0|\mu}E_{b,\nu}Q_{b,\nu}}{(P_{1|\mu}P_{0|\nu} - P_{1|\nu}P_{0|\mu})Y_{b,1}^L}.
\end{equation}

With all these quantities, Alice and Bob can compute the secret fraction
\begin{equation}
\begin{aligned}
    R & = \sum_k p_k P_{0|k} Y_{Z,0}^L
      + \sum_k p_k P_{1|k} Y_{Z,1}^L \left(1-h_2(e_{X,1}^U)\right) \\
    &-f\sum_k p_k Q_{Z,k} h_2(E_{Z,k}),
    \label{eq:rate}
\end{aligned}
\end{equation}

where $h_2$ is the binary entropy function and the last term considers the portion of key that is published in the error correction procedure, whose inefficiency is represented by $f$.
$R$ represents the fraction of detection events in basis $Z$ that Alice and Bob can consider secure.
Note how $e_{X,1}^U$ is calculated in basis $X$: this is because in the infinite-key scenario, the bit error in a basis converges to the phase error in the other.
The phase error in basis $Z$ is the quantity that measures the amount of information leaked to Eve and therefore should appear in Eq. \eqref{eq:rate}, but it is estimated by the bit error $e_{X,1}^U$ in basis $X$.

To conclude the analysis, we underline that the two restrictions \eqref{eq:first_restriction} and \eqref{eq:second_restriction} hold in common cases, such as Poisson, thermal and binomial distributions, when each of them is used as both ${\mathcal D}_\mu$ and ${\mathcal D}_\nu$, changing only the mean value. Moreover, the results achieved with this procedure comply with the distribution-specific ones present in the literature \cite{Ma2005,Curty2010}.

\section{Finite-key security bounds}\label{sec:finite}
We translate the above results into the finite-key scenario, in which Alice and Bob carry out their analysis on realistic key blocks of finite length and must consider statistical effects \cite{Tomamichel2012,Hayashi2014,Lim2014}.
In this case, it is more convenient to work with absolute numbers of events rather than conditional frequencies and probabilities. Hence, the quantities of interest are the number of detection events ($n_{b,k}$) and mismatches ($ m_{b,k}$) when both Alice and Bob chose the basis $b$ and Alice chose the probability distribution $\mathcal{D}_k$, and the number of detection events ($s_{b,i}$) and mismatches ($ v_{b,i}$) when Alice sent a pulse with a specific number of photons.

The first two are available experimentally, whereas the others must be estimated through upper and lower bounds.
Relations \eqref{eq:ExpValGain} and \eqref{eq:ExpValError} become:
\begin{align}
    \langle n_{b,k} \rangle &= \sum_i P_{k|i} s_{b,i},\\ 
    \langle m_{b,k} \rangle &= \sum_i P_{k|i} v_{b,i}.
\end{align}
Term $P_{k|i}$ should be found from $P_{i|k}$ using Bayes' theorem.
Due to finite statistics, we can no longer equate the expected values to the experimentally measured quantities.
To relate them, we can use Hoeffding's inequality and define confidence intervals \cite{Hoeffding1963}.
With probability at least $1-2\epsilon_{PE}$, we can find:
\begin{gather}
    \langle n_{b,k}\rangle \in[n_{b,k}^-, n_{b,k}^+]=[n_{b,k}-\delta_{nb}, n_{b,k}+\delta_{nb}]
\end{gather}
where $\delta_{nb}=\sqrt{(n_b/2)\ln(1/\epsilon_{PE})}$ and $n_b = \sum_k n_{b,k}$. An equivalent expression can be written for $\langle m_{b,k} \rangle$.

Then, the bounds of Sec. \ref{sec:asympt} for the vacuum+weak protocol can be translated into:
\begin{align}
    s_{b,0}^L &= \frac{\tau_0}{p_{\vac}} n_{b,\vac}^-, \\
    s_{b,0}^U &= \frac{\tau_0}{p_{\vac}} n_{b,\vac}^+, \\
    s_{b,1}^L &= \frac{\tau_1}{P_{1|\nu}-\alpha P_{1|\mu}} \cdot 
    \left(\frac{n_{b,\nu}^-}{p_\nu} - \alpha \frac{n_{b,\mu}^+}{p_\mu}-\frac{P_{0|\nu}-\alpha P_{0|\mu}}{\tau_0}s_{b,0}^U \right), \label{eq:finite_s1l_vw}\\
    v_{b,1}^U &= \min_{k\in\{\mu,\nu\}} \left( \frac{\tau_1}{P_{1|k}} \left(\frac{m_{b,k}^+}{p_k}-P_{0|k} \frac{m_{b,\vac}^-}{p_{\vac}} \right)\right).
\end{align}
where we have defined $\tau_i = \sum_k p_k P_{i|k}$ the probability that Alice sends an $i$-photon pulse.
In Eq. \eqref{eq:finite_s1l_vw}, $s_{b,0}^U$ should be replaced with $s_{b,0}^L$ if Eq. \eqref{eq:second_restriction} does not hold.

In the one-decoy case, these become:
\begin{align}
    s_{b,0}^L &= \frac{\tau_0}{P_{1|\mu}P_{0|\nu}-P_{1|\nu}P_{0|\mu}}\cdot
    \left(\frac{P_{1|\mu}}{p_\nu}n_{b,\nu}^- - \frac{P_{1|\nu}}{p_\mu}n_{b,\mu}^+\right), \\
    s_{b,0}^U &= \min_k\left(\frac{2m_{b,k}^+ \tau_0}{p_k P_{0|k}}\right) +2\delta_{nb}, \\
    s_{b,1}^L &= \frac{\tau_1}{P_{1|\nu}-\alpha P_{1|\mu}} \cdot 
    \left(\frac{n_{b,\nu}^-}{p_\nu} - \alpha \frac{n_{b,\mu}^+}{p_\mu}-\frac{P_{0|\nu}-\alpha P_{0|\mu}}{\tau_0}s_{b,0}^U \right), \\
    v_{b,1}^U &= \frac{\tau_1}{P_{1|\mu}P_{0|\nu}-P_{1|\nu}P_{0|\mu}}\cdot
    \left( \frac{P_{0|\nu}}{p_\mu} m_{b,\mu}^+ - \frac{P_{0|\mu}}{p_\nu} m_{b,\nu}^-\right).
\end{align}

Finally, we can no longer equate the phase error in basis $Z$ with the bit error in basis $X$, because this is true only if the latter is estimated with infinite statistics.
We find an upper bound on the phase error as \cite{Fung2010,Lim2014}:
\begin{equation}
    \phi_Z^U = \frac{v_{X,1}^U}{s_{X,1}^L} + \gamma\left(\epsilon_{PE}, \frac{v_{X,1}^U}{s_{X,1}^L}, s_{X,1}^L, s_{Z,1}^L\right) 
\end{equation}
where
\begin{equation}
    \gamma(a,b,c,d) = \sqrt{\frac{(c+d)(1-b)b}{cd\ln{2}} \cdot \log_2\left(\frac{c+d}{cd(1-b)ba^2}\right)}
\end{equation}

The length $\ell$ of the secret key that Alice and Bob can extract is:
\begin{equation}
\begin{aligned}
    \ell & =  s_{Z,0}^L + s_{Z,1}^L\cdot(1-h_2(\phi_Z^U)) \\
    & - f\sum_k n_{Z,k} h_2\left(\frac{m_{Z,k}}{n_{Z,k}}\right) \\
    & - 6\log_2\left(\frac{1}{\epsilon_{PE}}\right)-\log_2\left(\frac{2}{\epsilon_{hash}}\right).
    \label{eq:SKL}
\end{aligned}
\end{equation}

The last two terms, which have no equivalent in Eq. \eqref{eq:rate}, account for the bits that must be discarded for the secrecy analysis and confirmation of correctness.
In particular, $\epsilon_{hash}$ is the probability that non-identical keys pass the confirmation function.
It is common to choose $\epsilon_{PE}$ and $\epsilon_{hash}$ from the \textit{secrecy} and \textit{correctness} parameters $\epsilon_{sec}$ and $\epsilon_{cor}$ that Alice and Bob want to assign to the final keys.
Typical values for these are $\epsilon_{sec}= 10^{-9}$ and $\epsilon_{cor}= 10^{-15}$ \cite{Rusca2018}.
For our choice of protocols, we have $\epsilon_{sec}=18\epsilon_{PE}$ in the vacuum + weak case and $\epsilon_{sec}=19\epsilon_{PE}$ for one-decoy, whereas $\epsilon_{cor}= \epsilon_{hash}$ in both cases (the proof is similar to that of Ref. \cite{Lim2014}).

\section{Distribution-specific cases}\label{sec:simulations}

We use a realistic model of a QKD experiment to compare the performance of the two protocols (vacuum+weak and one-decoy) across two scenarios, roughly representing a high-end system (SNSPDs, GHz source, and low coding error) and a less expensive one (a single SPAD, slower source, and higher coding error).
We consider three different statistical distributions:
\begin{itemize}
    \item The Poisson distribution, characterized by its mean value $\mu$:
    \begin{equation}
        P^{(P)}_{i|\mu} = \frac{e^{-\mu}\mu^i}{i!}.
    \end{equation}
    This is the most common distribution in QKD, because it is produced with attenuated laser pulses.
    \item The thermal distribution, characterized by its mean value $\mu$:
    \begin{equation}
        P^{(T)}_{i|\mu} = \frac{\mu^i}{(\mu+1)^{i+1}}.
    \end{equation}
    This is the distribution of the number of photons in one arm of an unheralded SPDC source, and is also typical of classical incoherent light such as that produced by LEDs.
    \item The binomial distribution, characterized by its mean value $\mu$ and by the maximum number of photons $n$:
    \begin{equation}
        P^{(B)}_{i|\mu, n} = \binom{n}{i}\cdot \left(\frac{\mu}{n}\right)^i\cdot\left(1-\frac{\mu}{n}\right)^{n-i}.
    \end{equation}
    This distribution has never been used in QKD, but might become relevant in the near future with the development of new kinds of sources.
    For example, it describes the behavior of small collections of $n$ emitters that, when stimulated, might release exactly one photon, with a fixed probability $\mu/n$.
    If $n=1$, the decoy-state method is not even necessary, as photon-number-splitting is already excluded.
    However, if the emission probability is too small, it might be convenient to increase $n$ and use decoy states.
\end{itemize}

In all cases, we consider that Alice chooses between distributions of the same form, differing only by their mean value.
This is a practical choice, because the selection can be implemented rapidly with an intensity modulator which attenuates the output of a single optical source.
It can be shown that all three distributions maintain their form and change only in their mean value if attenuated.
Moreover, conditions \eqref{eq:first_restriction} and \eqref{eq:second_restriction} are met and for all three above considered distribution we have that $\alpha = \frac{P_{2|\nu}}{P_{2|\mu}}$.
We remark that, given Eq. \eqref{eq:second_restriction}, any implementation of the vacuum+weak protocol can in principle use the one-decoy bounds, however, to simplify the comparison, we do not allow this in our analysis, restricting each scheme to its own formulae.

For each of the two protocols, two scenarios and three distributions, and for a range of values of the global attenuation of an hypothetical QKD link, we optimize the protocol parameters ($p_Z$, $p_\mu$, $p_\nu$, $\mu$, $\nu$) to find the best attainable SKR using a simulated annealing algorithm \cite{Xiang1997}.
We keep the maximum number of photons for the binomial distribution fixed at $n=2$, and let $\mu$ reach this value in the optimization.
By doing this, we are considering an ideal situation where the $n$ emitters produce one photon each with unit probability, but the source can be attenuated to any $\mu \leq n$ if this improves the performance.
In all simulations, we keep the block size at $n_Z = 10^7$ bits and the security parameters at $\epsilon_{sec}= 10^{-9}$ and $\epsilon_{cor}= 10^{-15}$.

\begin{figure*}
    \centering
	\subfloat[High-end scenario, vacuum+weak protocol, binomial distribution.\label{fig:optimal_parameters_A}]{\includegraphics[width=0.98\columnwidth]{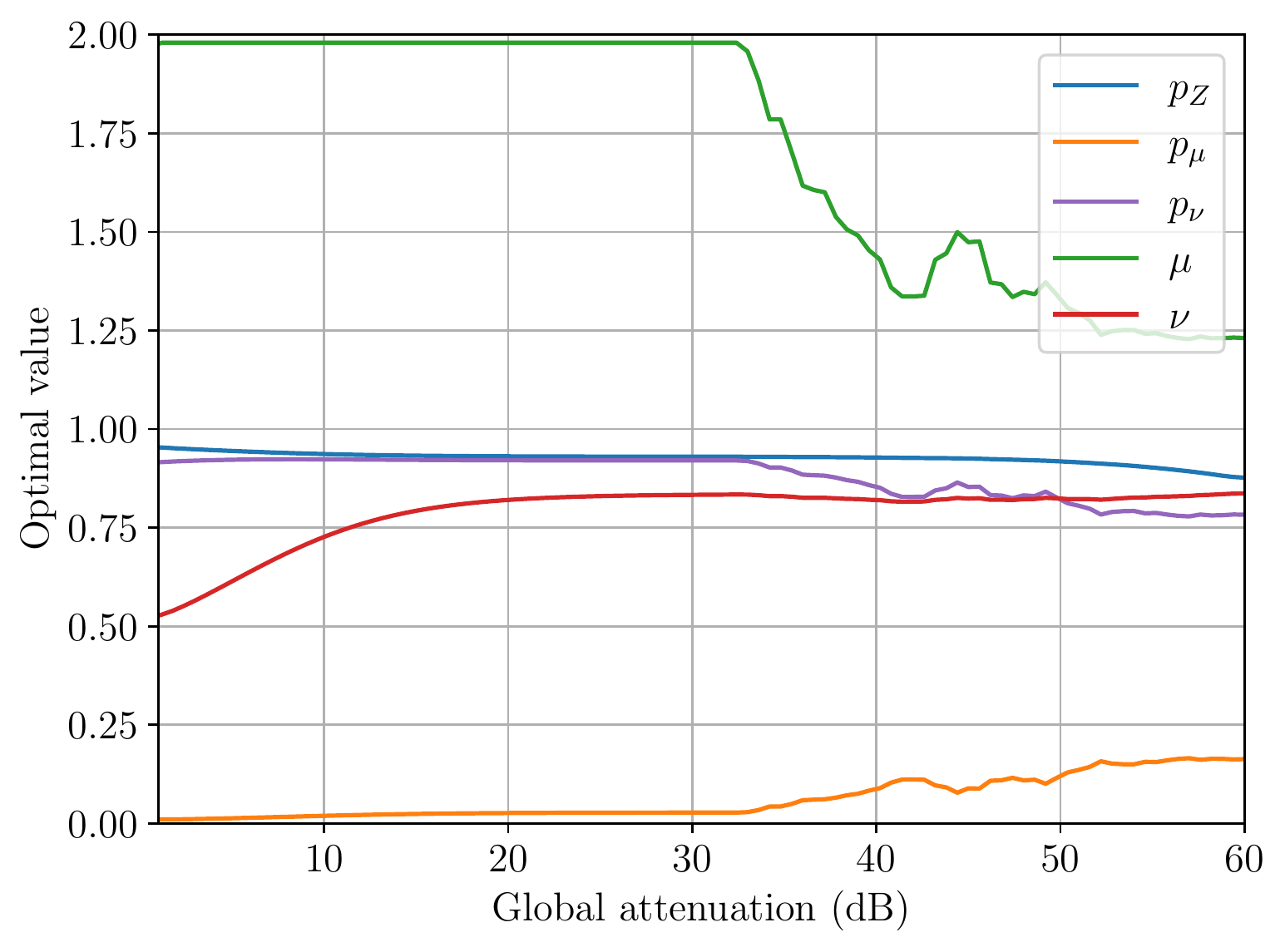}}
    \hfill
    \subfloat[Less expensive scenario, one-decoy protocol, thermal distribution.\label{fig:optimal_parameters_B}]{\includegraphics[width=0.98\columnwidth]{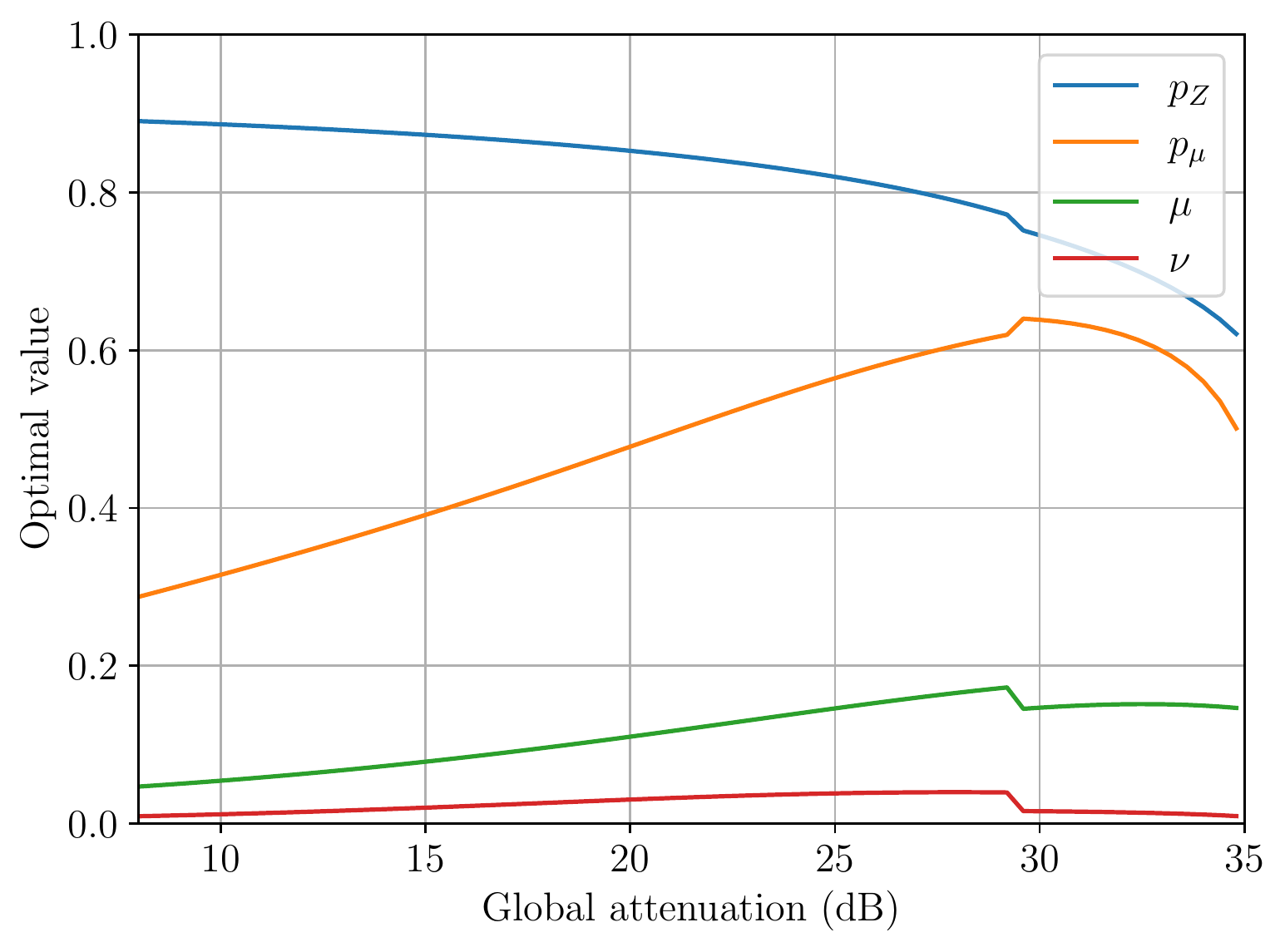}}
	\caption{Optimized protocol parameters.}
	\label{fig:optimal_parameters}
\end{figure*}

As an example, in Fig. \ref{fig:optimal_parameters} we show the results of the optimization for two of the twelve cases.
In Fig. \ref{fig:optimal_parameters_A}, we plot the results for the high-end scenario, with the vacuum+weak protocol and a binomial source.
Interestingly $\mu$ is larger than $1$ and even reaches the maximum value $n=2$ on the left side of the graph.
This can happen only for the binomial distribution, because even with such a high mean value, the number of photons in each pulse is upper bounded.
With such strong intensities, the parameter estimation is more accurate, and the key can still be built with the second level, which is chosen much more often.
The right side of the graph is slightly noisier because of the strong attenuation of the signal.
In this region, optimizing the protocol parameters has little effect and different values give similar results.

In Fig. \ref{fig:optimal_parameters_B}, we show the less expensive scenario, with the one-decoy protocol and thermal distribution.
The use of SPADs makes this system more prone to saturation and afterpulses, and hence the intensity levels stay low even for stronger losses.
The effect of saturation is especially visible in the ascending trend of $\mu$, $\nu$, and $p_\mu$: for weak attenuation, strong pulses cannot increase the raw key rate because the detectors are already saturated, and only increase the multi-photon emission probability; when losses grow, compensating them by increasing the average number of photons becomes a viable strategy.
The right side of the graph features a sharp jump: this is because there are several terms contributing to Eq. \eqref{eq:SKL}, and therefore several local maxima.
When one of them is promoted to global maximum, overcoming another, the optimal protocol parameters change.

\begin{figure*}
    \centering
	\subfloat[High-end scenario.]{\includegraphics[width=0.98\columnwidth]{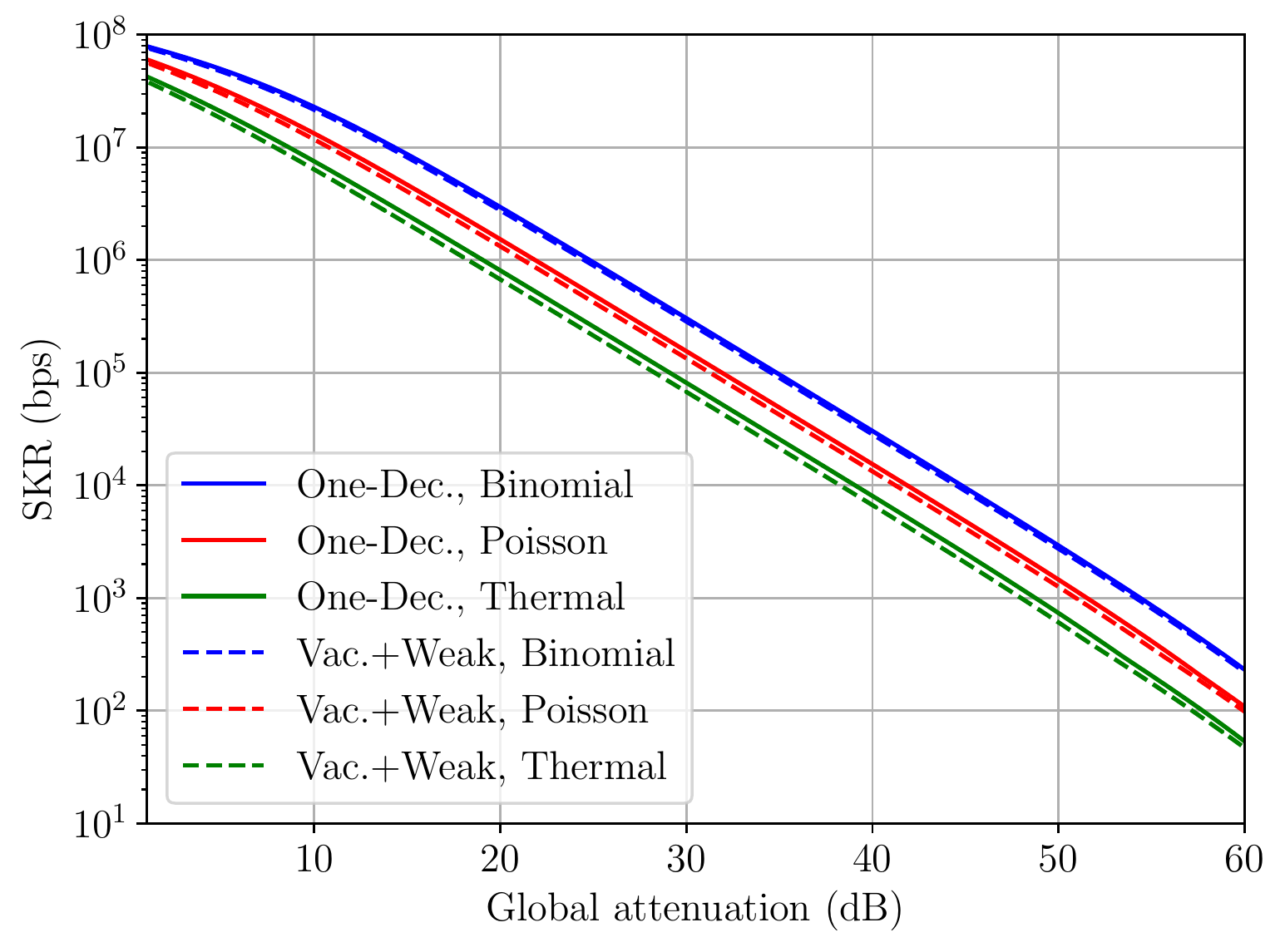}}
    \hfill
    \subfloat[Less expensive scenario.]{\includegraphics[width=0.98\columnwidth]{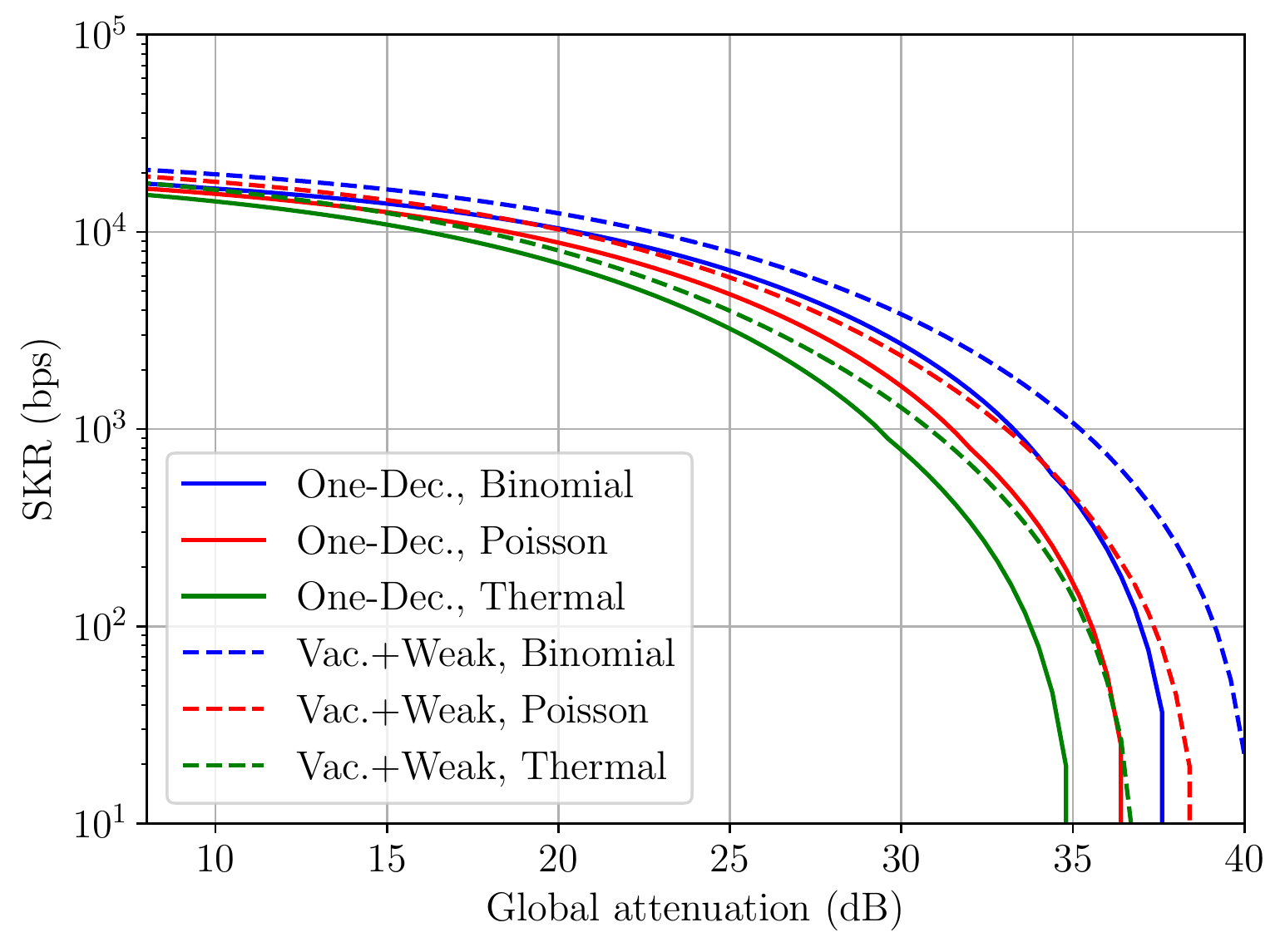}}
	\caption{Best attainable SKR using the optimized parameters for each scenario, protocol, statistical distribution and value of the global attenuation.}
	\label{fig:optimal_skr}
\end{figure*}

In Fig. \ref{fig:optimal_skr} we show the optimal SKR value.
All three distributions are close in terms of performance, confirming that the Poisson statistics is not the only reasonable choice for decoy states.
The thermal distribution performs the worst, due to its higher tails when the mean value is small, while the binomial one is the best, because it bounds the maximum number of emitted photons per pulse even if the mean intensities are high.

In the high-end scenario, the one-decoy protocol outperforms vacuum+weak.
This is a finite-key effect: the direct estimation of $s_{b,0}$ provided by the vacuum+weak protocol would require a higher $p_{\text{vac}}$ to accumulate more data, but that would reduce the signal rate and the overall SKR.
We can expect this to change if a longer block size is used ($\gtrsim 10^{10}$ bits).
In the less expensive scenario, the higher dark count rate of the detectors provides enough data for the vacuum+weak protocol, which becomes preferable.

As a further study of the binomial distribution, we optimize the parameters keeping fixed the global attenuation at 20 dB, and for different values of $n$ and $\max\left(\frac{\mu}{n}\right)\leq 1$, which represents the emission probability of a single photon by an individual emitter.
We do not fix $\frac{\mu}{n}$ at $\max\left(\frac{\mu}{n}\right)$ because it is always possible to attenuate the emission with optical elements.

\begin{figure*}
    \centering
	\subfloat[High-end scenario, vacuum+weak protocol.\label{fig:mnpsrA}]{\includegraphics[width=0.98\columnwidth]{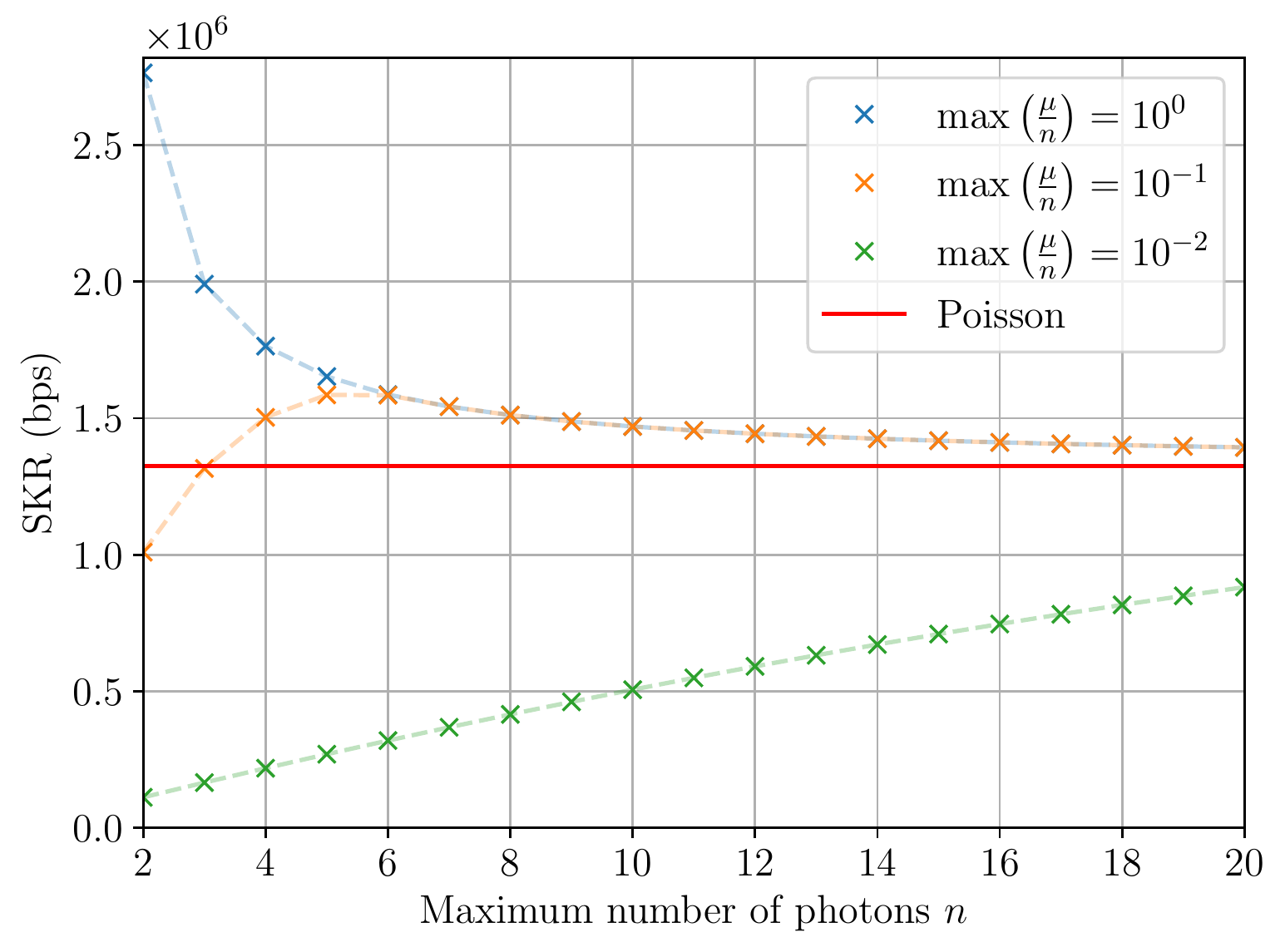}}
    \hfill
    \subfloat[Less expensive scenario, one-decoy protocol.\label{fig:mnpsrB}]{\includegraphics[width=0.98\columnwidth]{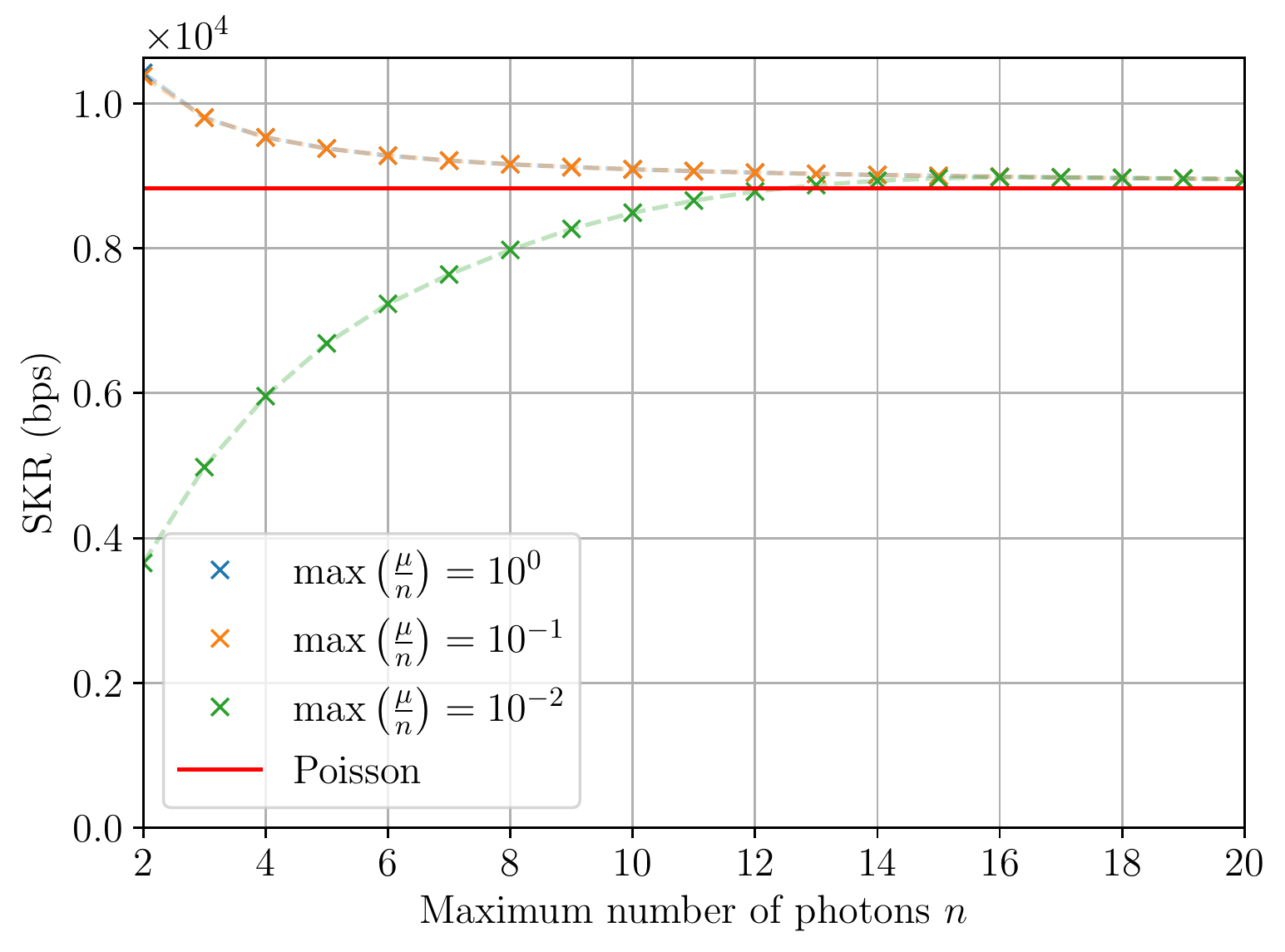}}
	\caption{Behavior of the SKR with the binomial distribution, for some fixed values of $\max\left(\frac{\mu}{n}\right)$.}
	\label{fig:mnpsr}
\end{figure*}

Figure \ref{fig:mnpsrA} shows the behavior of the SKR for the high-end scenario and vacuum+weak protocol.
While for $\max\left(\frac{\mu}{n}\right)=1$ adding more emitters is inconvenient because of the larger multi-photon probability, for $\max\left(\frac{\mu}{n}\right)=10^{-2}$, the SKR grows with $n$.
This means that more priority is given to increasing the detection rate, regardless of the multi-photon probability.
In the intermediate regime of $\max\left(\frac{\mu}{n}\right)=10^{-1}$, we see the SKR initially growing with $n$ and then decreasing after an optimum.
The right part of the curve coincides with that of $\max\left(\frac{\mu}{n}\right)=1$ because the same parameters are optimal in both cases.

Since the limit for large $n$ of the binomial distribution is the Poisson one, all three curves approach from above the performance obtained in the Poisson case.
This is more clear in Fig. \ref{fig:mnpsrB}, which reports the results of the less expensive scenario with the one-decoy protocol.
For any value of $\max\left(\frac{\mu}{n}\right)$, there is an optimal $n$ after which the performance decreases.
This can be explained considering that for large enough $n$ it is always possible to set $\mu$ and $\nu$ at the optimal values of the Poisson case.
Then, the two distributions are similar, with the binomial being slightly skewed towards lower numbers of photons, which increase the SKR.
The optimal $n$ is lower when the optimal $\mu$ of the Poisson case is lower, which happens in the second scenario because of afterpulses and saturation.
This is why we see the $\max\left(\frac{\mu}{n}\right)=10^{-2}$ curve overcome the Poisson line in Fig. \ref{fig:mnpsrB} but not in Fig. \ref{fig:mnpsrA}, for which a larger $n$ is needed.

\section{Conclusion}\label{sec:conclusion}
In this study, we showed that the decoy-state method, in its vacuum+weak and one-decoy forms, is  extendable to generic statistical distributions of the number of photons present in each QKD pulse.
We provided relations that consider finite-key effects and are directly applicable to experiments.

We evaluated the performance of the bounds for three relevant distributions, finding that thermal and binomial sources are comparable with standard coherent ones.
Specifically, the thermal distribution performs slightly worse due to its greater width, whereas the binomial one is the best because the maximum number of photons in each pulse is bounded.

With the development of new kinds of sources that are still vulnerable to the PNS attack but are not based on attenuated lasers, these results will provide a simple recipe for the optimal use of decoy states.

\begin{acknowledgments}
    This work was supported by: MIUR (Italian Minister for Education) under the initiative ``Departments of Excellence'' (Law 232/2016); 
     project "QUASIX" funded by the Italian Space Agency (ASI, Accordo n. 2019-5-U.0, CUP F44118000040001).
     CloudVeneto is acknowledged for the computational resources.
\end{acknowledgments}

\bibliographystyle{apsrev4-2}
\bibliography{bibliography}

\begin{thebibliography}{26}%
\makeatletter
\providecommand \@ifxundefined [1]{%
 \@ifx{#1\undefined}
}%
\providecommand \@ifnum [1]{%
 \ifnum #1\expandafter \@firstoftwo
 \else \expandafter \@secondoftwo
 \fi
}%
\providecommand \@ifx [1]{%
 \ifx #1\expandafter \@firstoftwo
 \else \expandafter \@secondoftwo
 \fi
}%
\providecommand \natexlab [1]{#1}%
\providecommand \enquote  [1]{``#1''}%
\providecommand \bibnamefont  [1]{#1}%
\providecommand \bibfnamefont [1]{#1}%
\providecommand \citenamefont [1]{#1}%
\providecommand \href@noop [0]{\@secondoftwo}%
\providecommand \href [0]{\begingroup \@sanitize@url \@href}%
\providecommand \@href[1]{\@@startlink{#1}\@@href}%
\providecommand \@@href[1]{\endgroup#1\@@endlink}%
\providecommand \@sanitize@url [0]{\catcode `\\12\catcode `\$12\catcode
  `\&12\catcode `\#12\catcode `\^12\catcode `\_12\catcode `\%12\relax}%
\providecommand \@@startlink[1]{}%
\providecommand \@@endlink[0]{}%
\providecommand \url  [0]{\begingroup\@sanitize@url \@url }%
\providecommand \@url [1]{\endgroup\@href {#1}{\urlprefix }}%
\providecommand \urlprefix  [0]{URL }%
\providecommand \Eprint [0]{\href }%
\providecommand \doibase [0]{https://doi.org/}%
\providecommand \selectlanguage [0]{\@gobble}%
\providecommand \bibinfo  [0]{\@secondoftwo}%
\providecommand \bibfield  [0]{\@secondoftwo}%
\providecommand \translation [1]{[#1]}%
\providecommand \BibitemOpen [0]{}%
\providecommand \bibitemStop [0]{}%
\providecommand \bibitemNoStop [0]{.\EOS\space}%
\providecommand \EOS [0]{\spacefactor3000\relax}%
\providecommand \BibitemShut  [1]{\csname bibitem#1\endcsname}%
\let\auto@bib@innerbib\@empty
\bibitem [{\citenamefont {Hwang}(2003)}]{Hwang2003}%
  \BibitemOpen
  \bibfield  {author} {\bibinfo {author} {\bibfnamefont {W.-Y.}\ \bibnamefont
  {Hwang}},\ }\href {https://doi.org/10.1103/PhysRevLett.91.057901} {\bibfield
  {journal} {\bibinfo  {journal} {Phys. Rev. Lett.}\ }\textbf {\bibinfo
  {volume} {91}},\ \bibinfo {pages} {057901} (\bibinfo {year}
  {2003})}\BibitemShut {NoStop}%
\bibitem [{\citenamefont {Liao}\ \emph {et~al.}(2017)\citenamefont {Liao},
  \citenamefont {Cai}, \citenamefont {Liu}, \citenamefont {Zhang},
  \citenamefont {Li}, \citenamefont {Ren}, \citenamefont {Yin}, \citenamefont
  {Shen}, \citenamefont {Cao}, \citenamefont {Li}, \citenamefont {Li},
  \citenamefont {Chen}, \citenamefont {Sun}, \citenamefont {Jia}, \citenamefont
  {Wu}, \citenamefont {Jiang}, \citenamefont {Wang}, \citenamefont {Huang},
  \citenamefont {Wang}, \citenamefont {Zhou}, \citenamefont {Deng},
  \citenamefont {Xi}, \citenamefont {Ma}, \citenamefont {Hu}, \citenamefont
  {Zhang}, \citenamefont {Chen}, \citenamefont {Liu}, \citenamefont {Wang},
  \citenamefont {Zhu}, \citenamefont {Lu}, \citenamefont {Shu}, \citenamefont
  {Peng}, \citenamefont {Wang},\ and\ \citenamefont {Pan}}]{Liao2017_Sat}%
  \BibitemOpen
  \bibfield  {author} {\bibinfo {author} {\bibfnamefont {S.-K.}\ \bibnamefont
  {Liao}}, \bibinfo {author} {\bibfnamefont {W.-Q.}\ \bibnamefont {Cai}},
  \bibinfo {author} {\bibfnamefont {W.-Y.}\ \bibnamefont {Liu}}, \bibinfo
  {author} {\bibfnamefont {L.}~\bibnamefont {Zhang}}, \bibinfo {author}
  {\bibfnamefont {Y.}~\bibnamefont {Li}}, \bibinfo {author} {\bibfnamefont
  {J.-G.}\ \bibnamefont {Ren}}, \bibinfo {author} {\bibfnamefont
  {J.}~\bibnamefont {Yin}}, \bibinfo {author} {\bibfnamefont {Q.}~\bibnamefont
  {Shen}}, \bibinfo {author} {\bibfnamefont {Y.}~\bibnamefont {Cao}}, \bibinfo
  {author} {\bibfnamefont {Z.-P.}\ \bibnamefont {Li}}, \bibinfo {author}
  {\bibfnamefont {F.-Z.}\ \bibnamefont {Li}}, \bibinfo {author} {\bibfnamefont
  {X.-W.}\ \bibnamefont {Chen}}, \bibinfo {author} {\bibfnamefont {L.-H.}\
  \bibnamefont {Sun}}, \bibinfo {author} {\bibfnamefont {J.-J.}\ \bibnamefont
  {Jia}}, \bibinfo {author} {\bibfnamefont {J.-C.}\ \bibnamefont {Wu}},
  \bibinfo {author} {\bibfnamefont {X.-J.}\ \bibnamefont {Jiang}}, \bibinfo
  {author} {\bibfnamefont {J.-F.}\ \bibnamefont {Wang}}, \bibinfo {author}
  {\bibfnamefont {Y.-M.}\ \bibnamefont {Huang}}, \bibinfo {author}
  {\bibfnamefont {Q.}~\bibnamefont {Wang}}, \bibinfo {author} {\bibfnamefont
  {Y.-L.}\ \bibnamefont {Zhou}}, \bibinfo {author} {\bibfnamefont
  {L.}~\bibnamefont {Deng}}, \bibinfo {author} {\bibfnamefont {T.}~\bibnamefont
  {Xi}}, \bibinfo {author} {\bibfnamefont {L.}~\bibnamefont {Ma}}, \bibinfo
  {author} {\bibfnamefont {T.}~\bibnamefont {Hu}}, \bibinfo {author}
  {\bibfnamefont {Q.}~\bibnamefont {Zhang}}, \bibinfo {author} {\bibfnamefont
  {Y.-A.}\ \bibnamefont {Chen}}, \bibinfo {author} {\bibfnamefont {N.-L.}\
  \bibnamefont {Liu}}, \bibinfo {author} {\bibfnamefont {X.-B.}\ \bibnamefont
  {Wang}}, \bibinfo {author} {\bibfnamefont {Z.-C.}\ \bibnamefont {Zhu}},
  \bibinfo {author} {\bibfnamefont {C.-Y.}\ \bibnamefont {Lu}}, \bibinfo
  {author} {\bibfnamefont {R.}~\bibnamefont {Shu}}, \bibinfo {author}
  {\bibfnamefont {C.-Z.}\ \bibnamefont {Peng}}, \bibinfo {author}
  {\bibfnamefont {J.-Y.}\ \bibnamefont {Wang}},\ and\ \bibinfo {author}
  {\bibfnamefont {J.-W.}\ \bibnamefont {Pan}},\ }\href
  {https://doi.org/10.1038/nature23655} {\bibfield  {journal} {\bibinfo
  {journal} {Nature}\ }\textbf {\bibinfo {volume} {549}},\ \bibinfo {pages}
  {43} (\bibinfo {year} {2017})}\BibitemShut {NoStop}%
\bibitem [{\citenamefont {Boaron}\ \emph {et~al.}(2018)\citenamefont {Boaron},
  \citenamefont {Boso}, \citenamefont {Rusca}, \citenamefont {Vulliez},
  \citenamefont {Autebert}, \citenamefont {Caloz}, \citenamefont {Perrenoud},
  \citenamefont {Gras}, \citenamefont {Bussi\`eres}, \citenamefont {Li},
  \citenamefont {Nolan}, \citenamefont {Martin},\ and\ \citenamefont
  {Zbinden}}]{Boaron2018}%
  \BibitemOpen
  \bibfield  {author} {\bibinfo {author} {\bibfnamefont {A.}~\bibnamefont
  {Boaron}}, \bibinfo {author} {\bibfnamefont {G.}~\bibnamefont {Boso}},
  \bibinfo {author} {\bibfnamefont {D.}~\bibnamefont {Rusca}}, \bibinfo
  {author} {\bibfnamefont {C.}~\bibnamefont {Vulliez}}, \bibinfo {author}
  {\bibfnamefont {C.}~\bibnamefont {Autebert}}, \bibinfo {author}
  {\bibfnamefont {M.}~\bibnamefont {Caloz}}, \bibinfo {author} {\bibfnamefont
  {M.}~\bibnamefont {Perrenoud}}, \bibinfo {author} {\bibfnamefont
  {G.}~\bibnamefont {Gras}}, \bibinfo {author} {\bibfnamefont {F.}~\bibnamefont
  {Bussi\`eres}}, \bibinfo {author} {\bibfnamefont {M.-J.}\ \bibnamefont {Li}},
  \bibinfo {author} {\bibfnamefont {D.}~\bibnamefont {Nolan}}, \bibinfo
  {author} {\bibfnamefont {A.}~\bibnamefont {Martin}},\ and\ \bibinfo {author}
  {\bibfnamefont {H.}~\bibnamefont {Zbinden}},\ }\href
  {https://doi.org/10.1103/PhysRevLett.121.190502} {\bibfield  {journal}
  {\bibinfo  {journal} {Phys. Rev. Lett.}\ }\textbf {\bibinfo {volume} {121}},\
  \bibinfo {pages} {190502} (\bibinfo {year} {2018})}\BibitemShut {NoStop}%
\bibitem [{\citenamefont {Dynes}\ \emph {et~al.}(2019)\citenamefont {Dynes},
  \citenamefont {Wonfor}, \citenamefont {Tam}, \citenamefont {Sharpe},
  \citenamefont {Takahashi}, \citenamefont {Lucamarini}, \citenamefont {Plews},
  \citenamefont {Yuan}, \citenamefont {Dixon}, \citenamefont {Cho},
  \citenamefont {Tanizawa}, \citenamefont {Elbers}, \citenamefont
  {Grei{\ss}er}, \citenamefont {White}, \citenamefont {Penty},\ and\
  \citenamefont {Shields}}]{Dynes2019}%
  \BibitemOpen
  \bibfield  {author} {\bibinfo {author} {\bibfnamefont {J.~F.}\ \bibnamefont
  {Dynes}}, \bibinfo {author} {\bibfnamefont {A.}~\bibnamefont {Wonfor}},
  \bibinfo {author} {\bibfnamefont {W.~W.-S.}\ \bibnamefont {Tam}}, \bibinfo
  {author} {\bibfnamefont {A.~W.}\ \bibnamefont {Sharpe}}, \bibinfo {author}
  {\bibfnamefont {R.}~\bibnamefont {Takahashi}}, \bibinfo {author}
  {\bibfnamefont {M.}~\bibnamefont {Lucamarini}}, \bibinfo {author}
  {\bibfnamefont {A.}~\bibnamefont {Plews}}, \bibinfo {author} {\bibfnamefont
  {Z.~L.}\ \bibnamefont {Yuan}}, \bibinfo {author} {\bibfnamefont {A.~R.}\
  \bibnamefont {Dixon}}, \bibinfo {author} {\bibfnamefont {J.}~\bibnamefont
  {Cho}}, \bibinfo {author} {\bibfnamefont {Y.}~\bibnamefont {Tanizawa}},
  \bibinfo {author} {\bibfnamefont {J.-P.}\ \bibnamefont {Elbers}}, \bibinfo
  {author} {\bibfnamefont {H.}~\bibnamefont {Grei{\ss}er}}, \bibinfo {author}
  {\bibfnamefont {I.~H.}\ \bibnamefont {White}}, \bibinfo {author}
  {\bibfnamefont {R.~V.}\ \bibnamefont {Penty}},\ and\ \bibinfo {author}
  {\bibfnamefont {A.~J.}\ \bibnamefont {Shields}},\ }\href
  {https://doi.org/10.1038/s41534-019-0221-4} {\bibfield  {journal} {\bibinfo
  {journal} {npj Quantum Inf.}\ }\textbf {\bibinfo {volume} {5}},\ \bibinfo
  {pages} {101} (\bibinfo {year} {2019})}\BibitemShut {NoStop}%
\bibitem [{\citenamefont {Avesani}\ \emph {et~al.}(2021)\citenamefont
  {Avesani}, \citenamefont {Calderaro}, \citenamefont {Foletto}, \citenamefont
  {Agnesi}, \citenamefont {Picciariello}, \citenamefont {Santagiustina},
  \citenamefont {Scriminich}, \citenamefont {Stanco}, \citenamefont {Vedovato},
  \citenamefont {Zahidy}, \citenamefont {Vallone},\ and\ \citenamefont
  {Villoresi}}]{Avesani:21}%
  \BibitemOpen
  \bibfield  {author} {\bibinfo {author} {\bibfnamefont {M.}~\bibnamefont
  {Avesani}}, \bibinfo {author} {\bibfnamefont {L.}~\bibnamefont {Calderaro}},
  \bibinfo {author} {\bibfnamefont {G.}~\bibnamefont {Foletto}}, \bibinfo
  {author} {\bibfnamefont {C.}~\bibnamefont {Agnesi}}, \bibinfo {author}
  {\bibfnamefont {F.}~\bibnamefont {Picciariello}}, \bibinfo {author}
  {\bibfnamefont {F.~B.~L.}\ \bibnamefont {Santagiustina}}, \bibinfo {author}
  {\bibfnamefont {A.}~\bibnamefont {Scriminich}}, \bibinfo {author}
  {\bibfnamefont {A.}~\bibnamefont {Stanco}}, \bibinfo {author} {\bibfnamefont
  {F.}~\bibnamefont {Vedovato}}, \bibinfo {author} {\bibfnamefont
  {M.}~\bibnamefont {Zahidy}}, \bibinfo {author} {\bibfnamefont
  {G.}~\bibnamefont {Vallone}},\ and\ \bibinfo {author} {\bibfnamefont
  {P.}~\bibnamefont {Villoresi}},\ }\href {https://doi.org/10.1364/OL.422890}
  {\bibfield  {journal} {\bibinfo  {journal} {Opt. Lett.}\ }\textbf {\bibinfo
  {volume} {46}},\ \bibinfo {pages} {2848} (\bibinfo {year}
  {2021})}\BibitemShut {NoStop}%
\bibitem [{\citenamefont {Brassard}\ \emph {et~al.}(2000)\citenamefont
  {Brassard}, \citenamefont {L\"utkenhaus}, \citenamefont {Mor},\ and\
  \citenamefont {Sanders}}]{PhysRevLett.85.1330}%
  \BibitemOpen
  \bibfield  {author} {\bibinfo {author} {\bibfnamefont {G.}~\bibnamefont
  {Brassard}}, \bibinfo {author} {\bibfnamefont {N.}~\bibnamefont
  {L\"utkenhaus}}, \bibinfo {author} {\bibfnamefont {T.}~\bibnamefont {Mor}},\
  and\ \bibinfo {author} {\bibfnamefont {B.~C.}\ \bibnamefont {Sanders}},\
  }\href {https://doi.org/10.1103/PhysRevLett.85.1330} {\bibfield  {journal}
  {\bibinfo  {journal} {Phys. Rev. Lett.}\ }\textbf {\bibinfo {volume} {85}},\
  \bibinfo {pages} {1330} (\bibinfo {year} {2000})}\BibitemShut {NoStop}%
\bibitem [{\citenamefont {Ma}\ \emph {et~al.}(2005)\citenamefont {Ma},
  \citenamefont {Qi}, \citenamefont {Zhao},\ and\ \citenamefont {Lo}}]{Ma2005}%
  \BibitemOpen
  \bibfield  {author} {\bibinfo {author} {\bibfnamefont {X.}~\bibnamefont
  {Ma}}, \bibinfo {author} {\bibfnamefont {B.}~\bibnamefont {Qi}}, \bibinfo
  {author} {\bibfnamefont {Y.}~\bibnamefont {Zhao}},\ and\ \bibinfo {author}
  {\bibfnamefont {H.-K.}\ \bibnamefont {Lo}},\ }\href
  {https://doi.org/10.1103/PhysRevA.72.012326} {\bibfield  {journal} {\bibinfo
  {journal} {Phys. Rev. A}\ }\textbf {\bibinfo {volume} {72}},\ \bibinfo
  {pages} {012326} (\bibinfo {year} {2005})}\BibitemShut {NoStop}%
\bibitem [{\citenamefont {Lim}\ \emph {et~al.}(2014)\citenamefont {Lim},
  \citenamefont {Curty}, \citenamefont {Walenta}, \citenamefont {Xu},\ and\
  \citenamefont {Zbinden}}]{Lim2014}%
  \BibitemOpen
  \bibfield  {author} {\bibinfo {author} {\bibfnamefont {C.~C.~W.}\
  \bibnamefont {Lim}}, \bibinfo {author} {\bibfnamefont {M.}~\bibnamefont
  {Curty}}, \bibinfo {author} {\bibfnamefont {N.}~\bibnamefont {Walenta}},
  \bibinfo {author} {\bibfnamefont {F.}~\bibnamefont {Xu}},\ and\ \bibinfo
  {author} {\bibfnamefont {H.}~\bibnamefont {Zbinden}},\ }\href
  {https://doi.org/10.1103/PhysRevA.89.022307} {\bibfield  {journal} {\bibinfo
  {journal} {Phys. Rev. A}\ }\textbf {\bibinfo {volume} {89}},\ \bibinfo
  {pages} {022307} (\bibinfo {year} {2014})}\BibitemShut {NoStop}%
\bibitem [{\citenamefont {Rusca}\ \emph {et~al.}(2018)\citenamefont {Rusca},
  \citenamefont {Boaron}, \citenamefont {Grünenfelder}, \citenamefont
  {Martin},\ and\ \citenamefont {Zbinden}}]{Rusca2018}%
  \BibitemOpen
  \bibfield  {author} {\bibinfo {author} {\bibfnamefont {D.}~\bibnamefont
  {Rusca}}, \bibinfo {author} {\bibfnamefont {A.}~\bibnamefont {Boaron}},
  \bibinfo {author} {\bibfnamefont {F.}~\bibnamefont {Grünenfelder}}, \bibinfo
  {author} {\bibfnamefont {A.}~\bibnamefont {Martin}},\ and\ \bibinfo {author}
  {\bibfnamefont {H.}~\bibnamefont {Zbinden}},\ }\href
  {https://doi.org/10.1063/1.5023340} {\bibfield  {journal} {\bibinfo
  {journal} {Appl. Phys. Lett.}\ }\textbf {\bibinfo {volume} {112}},\ \bibinfo
  {pages} {171104} (\bibinfo {year} {2018})}\BibitemShut {NoStop}%
\bibitem [{\citenamefont {Wang}\ \emph {et~al.}(2007)\citenamefont {Wang},
  \citenamefont {Wang},\ and\ \citenamefont {Guo}}]{Wang2007}%
  \BibitemOpen
  \bibfield  {author} {\bibinfo {author} {\bibfnamefont {Q.}~\bibnamefont
  {Wang}}, \bibinfo {author} {\bibfnamefont {X.-B.}\ \bibnamefont {Wang}},\
  and\ \bibinfo {author} {\bibfnamefont {G.-C.}\ \bibnamefont {Guo}},\ }\href
  {https://doi.org/10.1103/PhysRevA.75.012312} {\bibfield  {journal} {\bibinfo
  {journal} {Phys. Rev. A}\ }\textbf {\bibinfo {volume} {75}},\ \bibinfo
  {pages} {012312} (\bibinfo {year} {2007})}\BibitemShut {NoStop}%
\bibitem [{\citenamefont {Mauerer}\ and\ \citenamefont
  {Silberhorn}(2007)}]{Mauerer2007}%
  \BibitemOpen
  \bibfield  {author} {\bibinfo {author} {\bibfnamefont {W.}~\bibnamefont
  {Mauerer}}\ and\ \bibinfo {author} {\bibfnamefont {C.}~\bibnamefont
  {Silberhorn}},\ }\href {https://doi.org/10.1103/PhysRevA.75.050305}
  {\bibfield  {journal} {\bibinfo  {journal} {Phys. Rev. A}\ }\textbf {\bibinfo
  {volume} {75}},\ \bibinfo {pages} {050305} (\bibinfo {year}
  {2007})}\BibitemShut {NoStop}%
\bibitem [{\citenamefont {Adachi}\ \emph {et~al.}(2007)\citenamefont {Adachi},
  \citenamefont {Yamamoto}, \citenamefont {Koashi},\ and\ \citenamefont
  {Imoto}}]{Adachi2007}%
  \BibitemOpen
  \bibfield  {author} {\bibinfo {author} {\bibfnamefont {Y.}~\bibnamefont
  {Adachi}}, \bibinfo {author} {\bibfnamefont {T.}~\bibnamefont {Yamamoto}},
  \bibinfo {author} {\bibfnamefont {M.}~\bibnamefont {Koashi}},\ and\ \bibinfo
  {author} {\bibfnamefont {N.}~\bibnamefont {Imoto}},\ }\href
  {https://doi.org/10.1103/PhysRevLett.99.180503} {\bibfield  {journal}
  {\bibinfo  {journal} {Phys. Rev. Lett.}\ }\textbf {\bibinfo {volume} {99}},\
  \bibinfo {pages} {180503} (\bibinfo {year} {2007})}\BibitemShut {NoStop}%
\bibitem [{\citenamefont {Curty}\ \emph {et~al.}(2010)\citenamefont {Curty},
  \citenamefont {Ma}, \citenamefont {Qi},\ and\ \citenamefont
  {Moroder}}]{Curty2010}%
  \BibitemOpen
  \bibfield  {author} {\bibinfo {author} {\bibfnamefont {M.}~\bibnamefont
  {Curty}}, \bibinfo {author} {\bibfnamefont {X.}~\bibnamefont {Ma}}, \bibinfo
  {author} {\bibfnamefont {B.}~\bibnamefont {Qi}},\ and\ \bibinfo {author}
  {\bibfnamefont {T.}~\bibnamefont {Moroder}},\ }\href
  {https://doi.org/10.1103/PhysRevA.81.022310} {\bibfield  {journal} {\bibinfo
  {journal} {Phys. Rev. A}\ }\textbf {\bibinfo {volume} {81}},\ \bibinfo
  {pages} {022310} (\bibinfo {year} {2010})}\BibitemShut {NoStop}%
\bibitem [{\citenamefont {Huang}\ \emph {et~al.}(2018)\citenamefont {Huang},
  \citenamefont {Sun}, \citenamefont {Liu},\ and\ \citenamefont
  {Makarov}}]{Huang2018}%
  \BibitemOpen
  \bibfield  {author} {\bibinfo {author} {\bibfnamefont {A.}~\bibnamefont
  {Huang}}, \bibinfo {author} {\bibfnamefont {S.-H.}\ \bibnamefont {Sun}},
  \bibinfo {author} {\bibfnamefont {Z.}~\bibnamefont {Liu}},\ and\ \bibinfo
  {author} {\bibfnamefont {V.}~\bibnamefont {Makarov}},\ }\href
  {https://doi.org/10.1103/PhysRevA.98.012330} {\bibfield  {journal} {\bibinfo
  {journal} {Phys. Rev. A}\ }\textbf {\bibinfo {volume} {98}},\ \bibinfo
  {pages} {012330} (\bibinfo {year} {2018})}\BibitemShut {NoStop}%
\bibitem [{\citenamefont {Duligall}\ \emph {et~al.}(2006)\citenamefont
  {Duligall}, \citenamefont {Godfrey}, \citenamefont {Harrison}, \citenamefont
  {Munro},\ and\ \citenamefont {Rarity}}]{Duligall_2006}%
  \BibitemOpen
  \bibfield  {author} {\bibinfo {author} {\bibfnamefont {J.~L.}\ \bibnamefont
  {Duligall}}, \bibinfo {author} {\bibfnamefont {M.~S.}\ \bibnamefont
  {Godfrey}}, \bibinfo {author} {\bibfnamefont {K.~A.}\ \bibnamefont
  {Harrison}}, \bibinfo {author} {\bibfnamefont {W.~J.}\ \bibnamefont
  {Munro}},\ and\ \bibinfo {author} {\bibfnamefont {J.~G.}\ \bibnamefont
  {Rarity}},\ }\href {https://doi.org/10.1088/1367-2630/8/10/249} {\bibfield
  {journal} {\bibinfo  {journal} {New J. Phys.}\ }\textbf {\bibinfo {volume}
  {8}},\ \bibinfo {pages} {249} (\bibinfo {year} {2006})}\BibitemShut {NoStop}%
\bibitem [{\citenamefont {Xia}\ \emph {et~al.}(2019)\citenamefont {Xia},
  \citenamefont {Zhang}, \citenamefont {Xie}, \citenamefont {Yuan},
  \citenamefont {Lin}, \citenamefont {Liao}, \citenamefont {Liu}, \citenamefont
  {Peng}, \citenamefont {Zhang},\ and\ \citenamefont {Pan}}]{Xia:2019}%
  \BibitemOpen
  \bibfield  {author} {\bibinfo {author} {\bibfnamefont {X.-X.}\ \bibnamefont
  {Xia}}, \bibinfo {author} {\bibfnamefont {Z.}~\bibnamefont {Zhang}}, \bibinfo
  {author} {\bibfnamefont {H.-B.}\ \bibnamefont {Xie}}, \bibinfo {author}
  {\bibfnamefont {X.}~\bibnamefont {Yuan}}, \bibinfo {author} {\bibfnamefont
  {J.}~\bibnamefont {Lin}}, \bibinfo {author} {\bibfnamefont {S.-K.}\
  \bibnamefont {Liao}}, \bibinfo {author} {\bibfnamefont {Y.}~\bibnamefont
  {Liu}}, \bibinfo {author} {\bibfnamefont {C.-Z.}\ \bibnamefont {Peng}},
  \bibinfo {author} {\bibfnamefont {Q.}~\bibnamefont {Zhang}},\ and\ \bibinfo
  {author} {\bibfnamefont {J.-W.}\ \bibnamefont {Pan}},\ }\href
  {https://doi.org/10.1364/PRJ.7.001169} {\bibfield  {journal} {\bibinfo
  {journal} {Photon. Res.}\ }\textbf {\bibinfo {volume} {7}},\ \bibinfo {pages}
  {1169} (\bibinfo {year} {2019})}\BibitemShut {NoStop}%
\bibitem [{\citenamefont {Heindel}\ \emph {et~al.}(2012)\citenamefont
  {Heindel}, \citenamefont {Kessler}, \citenamefont {Rau}, \citenamefont
  {Schneider}, \citenamefont {Fürst}, \citenamefont {Hargart}, \citenamefont
  {Schulz}, \citenamefont {Eichfelder}, \citenamefont {Ro{\ss}bach},
  \citenamefont {Nauerth}, \citenamefont {Lermer}, \citenamefont {Weier},
  \citenamefont {Jetter}, \citenamefont {Kamp}, \citenamefont {Reitzenstein},
  \citenamefont {Höfling}, \citenamefont {Michler}, \citenamefont
  {Weinfurter},\ and\ \citenamefont {Forchel}}]{Heindel_2012}%
  \BibitemOpen
  \bibfield  {author} {\bibinfo {author} {\bibfnamefont {T.}~\bibnamefont
  {Heindel}}, \bibinfo {author} {\bibfnamefont {C.~A.}\ \bibnamefont
  {Kessler}}, \bibinfo {author} {\bibfnamefont {M.}~\bibnamefont {Rau}},
  \bibinfo {author} {\bibfnamefont {C.}~\bibnamefont {Schneider}}, \bibinfo
  {author} {\bibfnamefont {M.}~\bibnamefont {Fürst}}, \bibinfo {author}
  {\bibfnamefont {F.}~\bibnamefont {Hargart}}, \bibinfo {author} {\bibfnamefont
  {W.-M.}\ \bibnamefont {Schulz}}, \bibinfo {author} {\bibfnamefont
  {M.}~\bibnamefont {Eichfelder}}, \bibinfo {author} {\bibfnamefont
  {R.}~\bibnamefont {Ro{\ss}bach}}, \bibinfo {author} {\bibfnamefont
  {S.}~\bibnamefont {Nauerth}}, \bibinfo {author} {\bibfnamefont
  {M.}~\bibnamefont {Lermer}}, \bibinfo {author} {\bibfnamefont
  {H.}~\bibnamefont {Weier}}, \bibinfo {author} {\bibfnamefont
  {M.}~\bibnamefont {Jetter}}, \bibinfo {author} {\bibfnamefont
  {M.}~\bibnamefont {Kamp}}, \bibinfo {author} {\bibfnamefont {S.}~\bibnamefont
  {Reitzenstein}}, \bibinfo {author} {\bibfnamefont {S.}~\bibnamefont
  {Höfling}}, \bibinfo {author} {\bibfnamefont {P.}~\bibnamefont {Michler}},
  \bibinfo {author} {\bibfnamefont {H.}~\bibnamefont {Weinfurter}},\ and\
  \bibinfo {author} {\bibfnamefont {A.}~\bibnamefont {Forchel}},\ }\href
  {https://doi.org/10.1088/1367-2630/14/8/083001} {\bibfield  {journal}
  {\bibinfo  {journal} {New J. Phys.}\ }\textbf {\bibinfo {volume} {14}},\
  \bibinfo {pages} {083001} (\bibinfo {year} {2012})}\BibitemShut {NoStop}%
\bibitem [{\citenamefont {Takemoto}\ \emph {et~al.}(2015)\citenamefont
  {Takemoto}, \citenamefont {Nambu}, \citenamefont {Miyazawa}, \citenamefont
  {Sakuma}, \citenamefont {Yamamoto}, \citenamefont {Yorozu},\ and\
  \citenamefont {Arakawa}}]{Takemoto2015}%
  \BibitemOpen
  \bibfield  {author} {\bibinfo {author} {\bibfnamefont {K.}~\bibnamefont
  {Takemoto}}, \bibinfo {author} {\bibfnamefont {Y.}~\bibnamefont {Nambu}},
  \bibinfo {author} {\bibfnamefont {T.}~\bibnamefont {Miyazawa}}, \bibinfo
  {author} {\bibfnamefont {Y.}~\bibnamefont {Sakuma}}, \bibinfo {author}
  {\bibfnamefont {T.}~\bibnamefont {Yamamoto}}, \bibinfo {author}
  {\bibfnamefont {S.}~\bibnamefont {Yorozu}},\ and\ \bibinfo {author}
  {\bibfnamefont {Y.}~\bibnamefont {Arakawa}},\ }\href
  {https://doi.org/10.1038/srep14383} {\bibfield  {journal} {\bibinfo
  {journal} {Sci. Rep.}\ }\textbf {\bibinfo {volume} {5}},\ \bibinfo {pages}
  {14383} (\bibinfo {year} {2015})}\BibitemShut {NoStop}%
\bibitem [{\citenamefont {Lo}\ \emph {et~al.}(2005{\natexlab{a}})\citenamefont
  {Lo}, \citenamefont {Ma},\ and\ \citenamefont {Chen}}]{Lo2005}%
  \BibitemOpen
  \bibfield  {author} {\bibinfo {author} {\bibfnamefont {H.-K.}\ \bibnamefont
  {Lo}}, \bibinfo {author} {\bibfnamefont {X.}~\bibnamefont {Ma}},\ and\
  \bibinfo {author} {\bibfnamefont {K.}~\bibnamefont {Chen}},\ }\href
  {https://doi.org/10.1103/PhysRevLett.94.230504} {\bibfield  {journal}
  {\bibinfo  {journal} {Phys. Rev. Lett.}\ }\textbf {\bibinfo {volume} {94}},\
  \bibinfo {pages} {230504} (\bibinfo {year} {2005}{\natexlab{a}})}\BibitemShut
  {NoStop}%
\bibitem [{\citenamefont {Wang}(2005)}]{Wang2005}%
  \BibitemOpen
  \bibfield  {author} {\bibinfo {author} {\bibfnamefont {X.-B.}\ \bibnamefont
  {Wang}},\ }\href {https://doi.org/10.1103/PhysRevLett.94.230503} {\bibfield
  {journal} {\bibinfo  {journal} {Phys. Rev. Lett.}\ }\textbf {\bibinfo
  {volume} {94}},\ \bibinfo {pages} {230503} (\bibinfo {year}
  {2005})}\BibitemShut {NoStop}%
\bibitem [{\citenamefont {Lo}\ \emph {et~al.}(2005{\natexlab{b}})\citenamefont
  {Lo}, \citenamefont {Chau},\ and\ \citenamefont {Ardehali}}]{Lo2005a}%
  \BibitemOpen
  \bibfield  {author} {\bibinfo {author} {\bibfnamefont {H.-K.}\ \bibnamefont
  {Lo}}, \bibinfo {author} {\bibfnamefont {H.~F.}\ \bibnamefont {Chau}},\ and\
  \bibinfo {author} {\bibfnamefont {M.}~\bibnamefont {Ardehali}},\ }\href
  {https://doi.org/10.1007/s00145-004-0142-y} {\bibfield  {journal} {\bibinfo
  {journal} {J. Cryptol.}\ }\textbf {\bibinfo {volume} {18}},\ \bibinfo {pages}
  {133} (\bibinfo {year} {2005}{\natexlab{b}})}\BibitemShut {NoStop}%
\bibitem [{\citenamefont {Tomamichel}\ \emph {et~al.}(2012)\citenamefont
  {Tomamichel}, \citenamefont {Lim}, \citenamefont {Gisin},\ and\ \citenamefont
  {Renner}}]{Tomamichel2012}%
  \BibitemOpen
  \bibfield  {author} {\bibinfo {author} {\bibfnamefont {M.}~\bibnamefont
  {Tomamichel}}, \bibinfo {author} {\bibfnamefont {C.~C.~W.}\ \bibnamefont
  {Lim}}, \bibinfo {author} {\bibfnamefont {N.}~\bibnamefont {Gisin}},\ and\
  \bibinfo {author} {\bibfnamefont {R.}~\bibnamefont {Renner}},\ }\href
  {https://doi.org/10.1038/ncomms1631} {\bibfield  {journal} {\bibinfo
  {journal} {Nat. Commun.}\ }\textbf {\bibinfo {volume} {3}},\ \bibinfo {pages}
  {634} (\bibinfo {year} {2012})}\BibitemShut {NoStop}%
\bibitem [{\citenamefont {Hayashi}\ and\ \citenamefont
  {Nakayama}(2014)}]{Hayashi2014}%
  \BibitemOpen
  \bibfield  {author} {\bibinfo {author} {\bibfnamefont {M.}~\bibnamefont
  {Hayashi}}\ and\ \bibinfo {author} {\bibfnamefont {R.}~\bibnamefont
  {Nakayama}},\ }\href {https://doi.org/10.1088/1367-2630/16/6/063009}
  {\bibfield  {journal} {\bibinfo  {journal} {New J. Phys.}\ }\textbf {\bibinfo
  {volume} {16}},\ \bibinfo {pages} {063009} (\bibinfo {year}
  {2014})}\BibitemShut {NoStop}%
\bibitem [{\citenamefont {Hoeffding}(1963)}]{Hoeffding1963}%
  \BibitemOpen
  \bibfield  {author} {\bibinfo {author} {\bibfnamefont {W.}~\bibnamefont
  {Hoeffding}},\ }\href {https://doi.org/10.1080/01621459.1963.10500830}
  {\bibfield  {journal} {\bibinfo  {journal} {J. Am. Stat. Assoc.}\ }\textbf
  {\bibinfo {volume} {58}},\ \bibinfo {pages} {13} (\bibinfo {year}
  {1963})}\BibitemShut {NoStop}%
\bibitem [{\citenamefont {Fung}\ \emph {et~al.}(2010)\citenamefont {Fung},
  \citenamefont {Ma},\ and\ \citenamefont {Chau}}]{Fung2010}%
  \BibitemOpen
  \bibfield  {author} {\bibinfo {author} {\bibfnamefont {C.-H.~F.}\
  \bibnamefont {Fung}}, \bibinfo {author} {\bibfnamefont {X.}~\bibnamefont
  {Ma}},\ and\ \bibinfo {author} {\bibfnamefont {H.~F.}\ \bibnamefont {Chau}},\
  }\href {https://doi.org/10.1103/PhysRevA.81.012318} {\bibfield  {journal}
  {\bibinfo  {journal} {Phys. Rev. A}\ }\textbf {\bibinfo {volume} {81}},\
  \bibinfo {pages} {012318} (\bibinfo {year} {2010})}\BibitemShut {NoStop}%
\bibitem [{\citenamefont {Xiang}\ \emph {et~al.}(1997)\citenamefont {Xiang},
  \citenamefont {Sun}, \citenamefont {Fan},\ and\ \citenamefont
  {Gong}}]{Xiang1997}%
  \BibitemOpen
  \bibfield  {author} {\bibinfo {author} {\bibfnamefont {Y.}~\bibnamefont
  {Xiang}}, \bibinfo {author} {\bibfnamefont {D.}~\bibnamefont {Sun}}, \bibinfo
  {author} {\bibfnamefont {W.}~\bibnamefont {Fan}},\ and\ \bibinfo {author}
  {\bibfnamefont {X.}~\bibnamefont {Gong}},\ }\href
  {https://doi.org/10.1016/S0375-9601(97)00474-X} {\bibfield  {journal}
  {\bibinfo  {journal} {Phys. Lett. A}\ }\textbf {\bibinfo {volume} {233}},\
  \bibinfo {pages} {216} (\bibinfo {year} {1997})}\BibitemShut {NoStop}%
\end{thebibliography}%

\end{document}